\documentclass[twocolumn,amsmath,amssymb]{revtex4}
\usepackage{graphicx}
\usepackage{dcolumn}
\usepackage{bm}

\begin{document}

\title{Interpretation of Electron Tunneling from Uncertainty Principle}

\author{Angik Sarkar}
\affiliation{Department of Electrical Engineering,Indian Institute of Technology,Kharagpur,India}
\author{T.K.Bhattacharyya}
\email{tkb@ece.iitkgp.ernet.in}
\affiliation{Department of E and ECE,Indian Institute of Technology,Kharagpur,India}
\date{\today}

\begin{abstract}
Beginners studying quantum mechanics are often baffled with electron tunneling.Hence an easy approach for comprehension of the topic is presented here on the basis of uncertainty principle.An estimate of the tunneling time is also derived from the same method.
\end{abstract}
\maketitle
\section{Introduction}
The student who begins studying quantum mechanics is often at sea with the intricacies of the topic. The certainty with which physics predicted events in the classical domain is completely lacking here. Physics seems to have \textit{given up} and only seems contented with the calculation of probability of a particular event.\\
\indent The student is perhaps most intrigued by the quantum mechanical treatment of electron transmission across a potential barrier. It apparently contradicts the classical solution. Quantum mechanics predicts a probability of reflection even when the electron energy is higher than that of the barrier and a probability of transmission when the electron energy is less than that of the barrier.(The latter phenomenon is known as electron tunneling). However classical mechanics suggests that an object may cross a barrier only if its energy is higher than that of the barrier.For example,a ball will not cross a wall if its energy is lower than that required to cross the wall. The students wrest with these ideas for some time. But when they see that there is no way of avoiding the principles of quantum mechanics,they submit themselves to it and accept whatever it has to offer.\\
\indent The principle which the students seem to accept at the earliest is Heisenberg's Uncertainty Principle. Most of them, without a proper knowledge of higher mathematics, gape at the solutions that Schr$\ddot{o}$dinger's Equation has to offer. Without a proper visualization,the solutions seem vague to them.\\
\indent The purpose of this paper is to explain the phenomenon of tunneling on the basis of the Uncertainty Principle.However the approach is strictly pedagogical as no quantitative estimation may be given using it. It would only help the student appreciate the intricacies of quantum mechanics through a visualization of the phenomenon. Otherwise the beauty of the phenomenon would have remained buried in the solution of Schr$\ddot{o}$dinger's Equation.
\begin{figure}
\includegraphics[height = 9cm, width= 9cm]{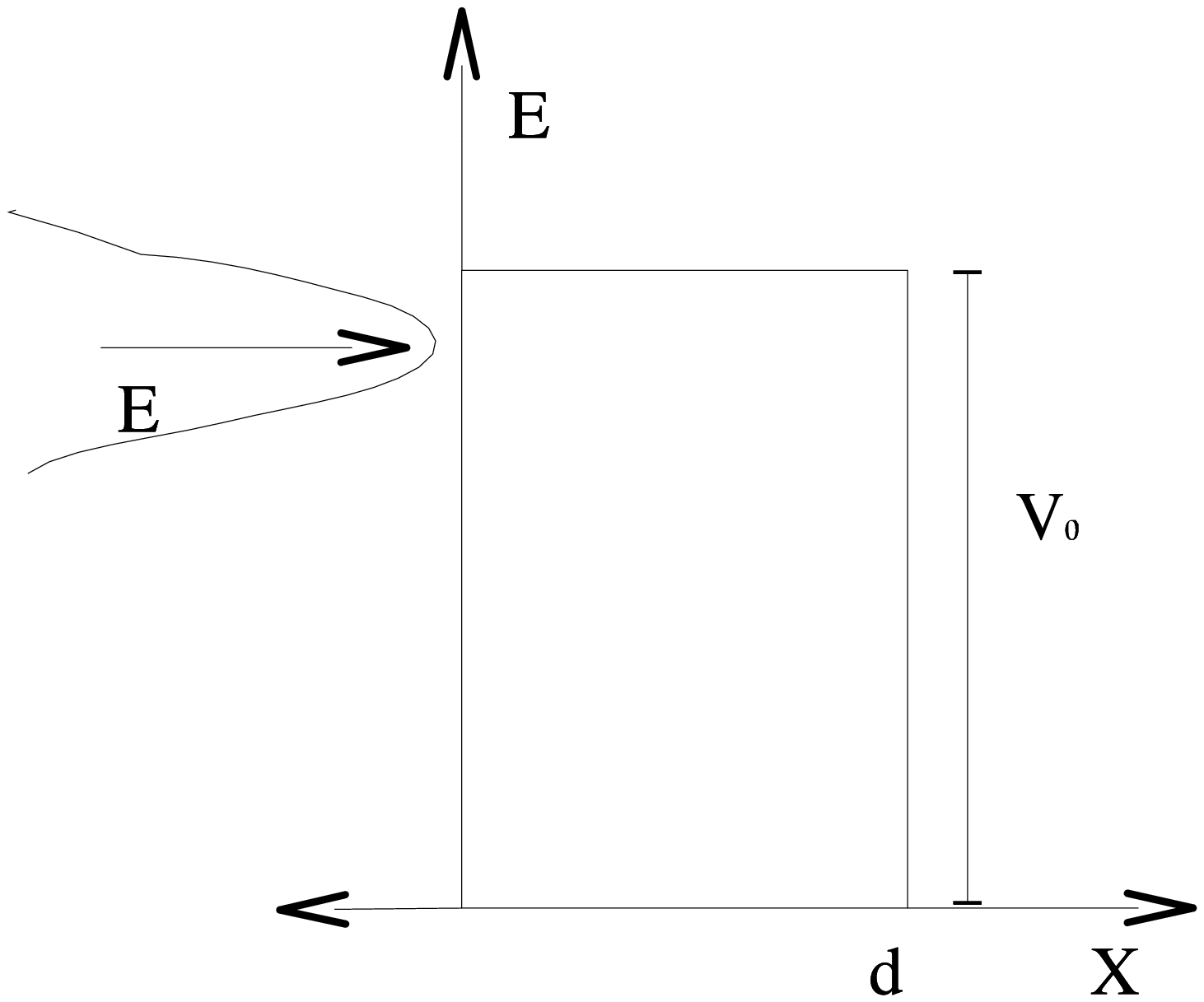}
\caption{\label{barrier}}The spread in electron energy of a packet incident at the barrier.Some of the electrons have energy higher than the barrier.
\end{figure}
\section{Schr\uppercase{$\ddot{o}$}dinger's Equation}
Normally the solution to the problem of transmission across a potential barrier is obtained by solving the time independent 1-D Schr$\ddot{o}$dinger's Equation.A rectangular potential barrier has been assumed for simplicity.(Refer to Fig. 1)
\begin{equation}
\begin{array}{lr}
\frac{d^2}{d{x^2}}\Psi{(x)}+\frac{2mE_{x}}{\hbar^2}\Psi{(x)}=0& \textbf{x}<0\quad\&\quad\textbf{x}>\textbf{d}\\
\frac{{d^2}}{d{x^2}}\Psi{(x)}+\frac{2m}{\hbar^2}[E_{x}-V_0]\Psi{(x)}=0&\textbf{x}>0\quad\&\quad\textbf{x}<\textbf{d}
\end{array}
\end{equation}
As given in any standard book on quantum mechanics\cite{quan}, there are two solution to this problem-one for $E > V_0 $ ,another for $E < V_0 $. The solutions are as given below.\\
\begin{center}
\underline{\textit{Case 1} : $E > V_0 $}
\end{center}
\begin{equation}
\begin{array}{lr}

\Psi(x) = A\exp(ikx) + B\exp(-ikx) & \textbf{x}<0\\
k = \frac{\sqrt{2mE}}{\hbar}&\\

\Psi(x) = C\exp(ik_1x) + D\exp(-ik_1x) & \textbf{x}>0\quad\&\quad\textbf{x}< d\\
k_1 = \frac{\sqrt{2m(E-V_0)}}{\hbar}&\\

\Psi(x) = E\exp(ikx)& \textbf{x}> d\\

\end{array}
\end{equation}
 
\begin{center}
\underline{\textit{Case 2} : $E < V_0 $}
\end{center}
\begin{equation}
\begin{array}{lr}

\Psi(x) = A\exp(ikx) + B\exp(-ikx) & \textbf{x}<0\\
k = \frac{\sqrt{2mE}}{\hbar}&\\

\Psi(x) = C\exp(\alpha x) + D\exp(-\alpha x) & \textbf{x}>0\quad\&\quad\textbf{x}< d\\
\alpha = \frac{\sqrt{2m(V_0-E)}}{\hbar}&\\

\Psi(x) = E\exp(ikx)& \textbf{x}> d\\

\end{array}
\end{equation}

Here A,B,C,D,E are constants.\\ 
\indent From the solutions it is proved mathematically that there is a definite probability that an electron with energy higher than that of the barrier might be reflected.Also from the solution for the case $ E < V_0 $, there is a probability for these electrons to be transmitted. These solutions, despite giving a quantitative estimate of the probability fail to provide an intuitive understanding of the phenomena.

\section{Uncertainty Principle}
The first principle of quantum mechanics that a student is exposed is probably the Heisenberg's Uncertainty Principle.In the most general terms,it states that if an electron is in state $|\Psi\rangle$ and A and B are two observables,then
\begin{equation}
\Delta(A)\Delta(B)\geq \frac{\left|\langle \Psi |[A,B]|\Psi\rangle\right|}{2}\label{uncer}
\end{equation}
The two forms of this principle which the students are most familiar with are:
\begin{eqnarray}
\Delta{E}\Delta{t}\geq \hbar\label{E}\\
\Delta{p}\Delta{x}\geq \hbar 
\end{eqnarray}
E is the energy of the electron; p is the momentum of the electron in the x direction.\\\indent Now we can apply this principle to the case of transmission across a potential barrier.
Let a packet of electrons be incident at a potential barrier of height $V_0$. By Uncertainty Principle we know the position of the electrons with an accuracy of $\Delta{x}$(say). Now if the $\Delta{x}$ is greater than the width of the barrier, then there is a definite probability of some electrons being on the other side of the barrier. This case is investigated more closely. For a standard potential barrier,let the width of the barrier be d. Let $\Delta{x}$ be equal to 1 nm. Calculating, $\Delta{p}$ for this case approximately turns out to be $1.054e^{-25}$ Kg m/s. Energy of an electron may be related to its momentum in the familiar classical form as, 
\begin{equation}
E=\frac{p^2}{2m^{*}}
\end{equation} 
m is the mass of the free electron.
$m^{*}$ is the effective mass of the electron.
Hence,
\begin{equation}
\Delta{E}=\frac{p\Delta{p}}{m^{*}}
\end{equation}
The momentum p is very small,we can safely assume,
\begin{eqnarray}
p&\approx &\Delta{p}\\
\Delta{E}& = &\frac{(\Delta{p})^2}{m^{*}}
\end{eqnarray}
Substituting the value of $\Delta{p}$ and assuming $m^{*}$ to be 0.07m(This is frequently encountered in solid state physics),we can calculate $\Delta{E}$ to be approximately 1.1 eV.
Now this value of uncertainty in energy is comparable to the value of barrier heights frequently encountered.(The barrier height for Si-Si$0_2$ interface in CMOS is roughly 3.1 eV)
Substituting this value of $\Delta{E}$ in (\ref{E}) ,we can get $\Delta{t}$ to be of the order of $10^{-15}$ s.

\section{Inference}
In the previous section some approximate calculations have been made. In this section the physics buried in those calculations will be unfurled. Let us make an assumption that the electrons will be transmitted across the barrier only if their individual energy is greater than the height of the barrier($V_0$) i.e. condition for transmission is same as that in classical physics. When electrons are incident on the barrier there is a spread in energy of the electrons.As the uncertainty in position gets smaller,the uncertainty in momentum and energy increases. For barriers small enough, the uncertainty in energy is comparable to the height of the barrier.Thus there is a definite probability that some of the electrons in a group with energy E($E<V_O$) may possess energy which is higher than that of the barrier.Thus these electrons may cross the barrier while the others may not. Similarly in the case where $E>V_0$, there is a definite probability that there may be electrons which possess energy less than that of the barrier.According to our assumption,there would be some electrons which would be reflected.Hence using our classical assumption we have accounted for all the events occurring in various cases of transmission of electrons across a potential barrier. All the events are equivalent to the classical case.Thus actually tunneling is not a phenomenon where some classically forbidden event takes place. It is only that due the uncertainty in the observables, there is a definite probability for the apparent non-classical events to occur.\\ 
        \indent In (\ref{E}), $\Delta{t}$ is the uncertainty of time over which the measurement of energy has been made.The energy measurement is done on the either side of the barrier. Thus the uncertainty of time is approximately equal to the transit time of electrons through the barrier. In the previous section the uncertainty in time(equal to tunneling time) was calculated to be of the order of $10^{-15}$ s. This is in excellent agreement with the detailed calculations performed recently\cite{Stein,but,Chiao,St,Dav}.\\
        \indent One might wonder where does the uncertainty in electron energy come from? The answer to this probably lies in the inelastic scattering at the incident surface. The electrons might absorb energy from the lattice(through inelastic processes) and rise to higher energies.
        
\section{Conclusion}
A qualitative explanation of electron transmission through a potential barrier has been presented here.This explanation is based on the Uncertainty Principle.The explanation has also been justified with a sample calculation of uncertainties.The explanation will be most useful for explaining the essence of tunneling to novices. Most textbooks present the tunneling effect as one of the phenomena for which the difference between classical and quantum physics is most striking. This leads to confusion in the students. The explanation provided here is amicable to the classical perceptions. Thus it is expected that the students will accept this explanation easily. Moreover, this also illustrates the probabilistic nature of quantum nature in a better way.\\
\indent Many students initially exposed to Uncertainty Principle often tend to interpret (\ref{uncer}) in the following way: measuring an observable A to some accuracy $\Delta{A}$ causes the value of B to be 'disturbed' by an amount $\Delta{B}$ in such a way that some sort of inequality similar to (\ref{uncer}) is being satisfied.While it is true that measurements cause disturbance to the system being measured,this is not the content of uncertainty principle. The correct interpretation is that if we prepare a large number of quantum systems in identical states,$|\Psi\rangle$, and then perform measurements of A on some of those systems and B in others then the standard deviation $\Delta{A}$ times the standard deviation $\Delta{B}$ will satisfy (\ref{uncer}). The explanation given in this paper highlights this interpretation. Hence the explanation will also help students understand the Uncertainty Principle in a better way.\\
\indent The explanation requires a bare minimum of mathematics. It provides a valuable insight into the governing mechanisms which would have remained buried in the mathematical complexities of the solutions. Thus we hope that it would be useful in teaching of undergraduate quantum mechanics and modern physics.

\end{document}